\documentclass[12pt]{article}
\usepackage{epsfig}
\usepackage{amssymb}
\usepackage{amsmath}
\usepackage{amsfonts}
\usepackage{graphicx}
\usepackage{mathrsfs}
\usepackage[dvips]{color}
\usepackage{multirow}

\newcommand{\R}{\mathbb{R}}
\newcommand{\C}{\mathbb{C}}

\newcommand{\fa}{\mathfrak{a}}

\newcommand{\fm}{\mathfrak{m}}
\newcommand{\fn}{{\,\mathfrak{n}\,}}

\newcommand{\bA}{\mathbf{A}}

\newcommand{\bH}{\mathbf{H}}
\newcommand{\bI}{\mathbf{I}}

\newcommand{\bM}{\mathbf{M}}
\newcommand{\bN}{\mathbf{N}}

\newcommand{\bS}{\mathbf{S}}

\newcommand{\bU}{\mathbf{U}}

\newcommand{\bsigma}{\mathbf{\sigma}}

\newcommand{\cO}{\mathcal{O}}

\newcommand{\cS}{\mathcal{S}}

\newcommand{\cU}{\mathcal{U}}

\newcommand{\be}{\begin{equation}}
\newcommand{\ee}{\end{equation}}
\newcommand{\bea}{\begin{eqnarray}}
\newcommand{\eea}{\end{eqnarray}}
\newcommand{\nn}{\nonumber}
\newcommand{\kt}{\rangle}
\newcommand{\br}{\langle}

\newcommand{\ed}{\end{document}}

\newcommand{\bi}{\begin{itemize}}
\newcommand{\ei}{\end{itemize}}

\newcommand{\bce}{\begin{center}}
\newcommand{\ece}{\end{center}}

\newcommand{\sT}{\mathscr{T}}

\begin{document}

\title{Adiabatic Series Expansion and Higher-Order Semiclassical Approximations in Scattering Theory}

\author{Ali~Mostafazadeh\thanks{E-mail address:
amostafazadeh@ku.edu.tr, Phone: +90 212 338 1462, Fax: +90 212 338
1559}
\\
Department of Mathematics, Ko\c{c} University,\\
34450 Sar{\i}yer, Istanbul, Turkey}
\date{ }
\maketitle

\begin{abstract}

The scattering properties of any complex scattering potential, $v:\R\to\C$, can be obtained from the dynamics of a particular non-unitary two-level quantum system ${\cS}_v$. The application of the adiabatic approximation to ${\cS}_v$ yields a semiclassical treatment of the scattering problem. We examine the adiabatic series expansion for the evolution operator of ${\cS}_v$ and use it to obtain corrections of arbitrary order to the semiclassical formula for the transfer matrix of $v$. This results in a high-energy approximation scheme that unlike the semiclassical approximation can be applied for potentials with large derivatives.

\medskip

\end{abstract}

\maketitle

Complex scattering potentials \cite{muga} are capable of supporting intriguing phenomena such as spectral singularities \cite{prl-2009,ss,jpa-2012}, unidirectional reflectionlessness \cite{ur}, and unidirectional invisibility \cite{invisible,invisible2,pra-2013a}. Recently, we showed that the transfer matrix $\bM$ of a general scattering potential $v(x)$ could be identified with the $S$-matrix of a non-unitary two-level quantum system ${\cS}_v$, \cite{pra-2014}. This reduces the scattering problem for $v$ to the determination of the evolution operator $\bU(\tau,\tau_0)$ of ${\cS}_v$. In Ref.~\cite{p115}, we explore the application of the adiabatic approximation to determine $\bU(\tau,\tau_0)$ and show that this gives rise to a semiclassical treatment of the scattering problem. In the present article, we conduct a detailed study of an improved adiabatic (or semiclassical) approximation scheme for the computation $\bM$ that is based on the adiabatic series expansion of $\bU(\tau,\tau_0)$, \cite{pra-1997a,jmp-1999,book1}.

The Hamiltonian for the system ${\cS}_v$ has the form \cite{pra-2014,ap-2014}
    \begin{align}
	\bH(\tau)&:=\left[\begin{array}{cc}
	w(\tau)-1 & w(\tau)\\
	-w(\tau) & -w(\tau)+1\end{array}\right]
	=-\bsigma_3+w(\tau)\bN,
	\label{def2}\\
	w(\tau)&:=\frac{v(\tau/k)}{2k^2},~~~~~~~~~
    \bN:=i\bsigma_2+\bsigma_3=
    \left[\begin{array}{cc} 1 & 1 \\-1 & -1\end{array}\right],
	\end{align}
where $\tau:=kx$, $k$ is the wavenumber, and $\bsigma_i$, with $i=1,2,3$, are Pauli matrices. The $S$-matrix of $\bH(\tau)$, that we denote by $\bS$, satisfies \cite{pra-2014}
    \be
    \bS:=\bU_0(+\infty)^{\!-1}\,\bU(+\infty,-\infty)\bU_0(-\infty)=
    \bU_0(\tau_+)^{\!-1}\,\bU(\tau_+,\tau_-)\bU_0(\tau_-)=\bM,
    \label{S}
    \ee
where $(\tau_-,\tau_+)$ is the largest open interval outside which $v(x)$ vanishes identically,
$\bU_0(\tau):=e^{i\tau\sigma_3}$ is the evolution operator for the Hamiltonian $\bH_0:=-\sigma_3$, which corresponds to setting $v=0$ in $\bH(\tau)$, $\bU(\tau,\tau_0):=\sT\exp\left\{-i\int_{\tau_0}^\tau\bH(\tau')d\tau'\right\}$ is the evolution operator for $\bH(\tau)$, and $\sT$ is the time-ordering operator.

Let us also recall that the transfer matrix \cite{sanchez-soto} is related to the reflection amplitudes $R^{r/l}$ from right/left and the transmission amplitude $T$ according to \cite{jpa-2012,pra-2013a}
    \be
    \bM=\left[\begin{array}{cc}
    \displaystyle T-\frac{R^lR^r}{T} & \displaystyle \frac{R^r}{T} \vspace{.2cm}\\
    \displaystyle -\frac{R^l}{T} & \displaystyle \frac{1}{T}\end{array}\right].
    \label{M}
    \ee
In view of (\ref{S}) and (\ref{M}), the determination of the evolution operator $\bU(\tau,\tau_0)$ gives a complete solution of the scattering problem defined by the potential $v$.

In general, it is not possible to compute $\bU(\tau,\tau_0)$ exactly. This motivates the use of the adiabatic approximation, which we can conveniently describe by \cite{p115}
    \be
    \bU(\tau,\tau_0)\approx \cU_0(\tau,\tau_0):=\sum_{a=\pm} e^{i([\delta_a(\tau)+\gamma_a(\tau)]}|\Psi_a(\tau)\kt\br\Phi_a(\tau_0)|.
    \label{adia-U}
    \ee
Here
    \begin{align}
    &\delta_\pm(\tau):=-\int_{\tau_0}^\tau E_\pm(\tau')d\tau', && 
    \gamma_\pm(\tau):=i\int_{\tau_0}^\tau \br\Phi_\pm(\tau')|d\Psi_\pm(\tau')\kt d\tau',
    \label{d-phi}\\
    & E_\pm(\tau):=\pm\fn(\tau),
    && \fn(\tau):=\sqrt{1- 2 w(\tau)}=\sqrt{1-\frac{v(\tau/k)}{k^2}},
    \label{d-E=}\\
    &\Psi_\pm(\tau):=\frac{1}{2}\left[\begin{array}{c}
	1\mp\fn(\tau)\\
	1\pm\fn(\tau)\end{array}\right],
    &&\Phi_\pm(\tau):=\frac{1}{2\fn(\tau)^*}\left[\begin{array}{c}
    \fn(\tau)^*\mp 1\\
    \fn(\tau)^*\pm 1\end{array}\right].
	\label{eg-val-vec}
	\end{align}
Note that $\Psi_\pm$ and $\Phi_\pm$ are respectively eigenvectors of $\bH(\tau)$ and $\bH(\tau)^\dagger$ with eigenvalue $E_\pm$ and $E_\pm^*$. They form a complete biorthonormal system \cite{review}, i.e.,
    \be
    \br \Phi_a|\Psi_b\kt=\delta_{ab},~~~~~~~~~~~~
    \sum_{a=1}^2|\Psi_a\kt\br \Phi_a|=\bI,
    \label{cbos}
    \ee
where $\delta_{ab}$ stands for the Kronecker delta symbol and $\bI$ is the $2\times 2$ identity matrix. The $\delta_\pm(\tau)$ and $\gamma_\pm(\tau)$ appearing in (\ref{adia-U}) are respectively the dynamical and geometric parts of the phase of the adiabatically evolving state vectors having $\Psi_\pm(\tau_0)$ as their initial value, \cite{GW}.

In view of (\ref{d-phi}) -- (\ref{eg-val-vec}), we have \cite{p115}
    \be
    e^{i\gamma_\pm(\tau)}=\sqrt{\frac{\fn(\tau_0)}{\fn(\tau)}}.
    \label{geo=}
    \ee
It is also easy to show that the adiabatic approximation (\ref{adia-U}) is reliable provided that
    \be
	\left| \frac{\dot\fn(\tau)}{4\fn(\tau)^2}\right|\ll 1.
	\label{adi-condi-0}
	\ee
As noted in Ref.~\cite{p115}, this is equivalent to the condition for the validity of the semiclassical approximation, namely $|v'(x)|/8|k^2-v(x)|^{3/2}\ll 1$.

In Refs.~\cite{pra-1997a,book1} we outline a generalization of adiabatic approximation that corresponds to truncations of the adiabatic series expansion for the evolution operator. In the following sections we examine the application of this method for the calculation of $\bU(\tau,\tau_0)$ and subsequently $\bM$. Because the adiabatic approximation~(\ref{adia-U}) leads to a semiclassical expression for $\bM$, the above generalized adiabatic approximation scheme gives rise to a hierarchy of improved semiclassical approximations in one-dimensional potential scattering.

Consider making a time-dependent linear transformation generated by the inverse of the adiabatic time-evolution operator \cite{pra-1997a,jmp-1999,book1}, i.e., $\Psi(\tau)\to \tilde\Psi(\tau):= \cU_0(\tau,\tau_0)^{-1}\Psi(\tau)$. Under this transformation the Hamiltonian $\bH(\tau)$ and the time-evolution operator $\bU(\tau,\tau_0)$ transform according to
    \bea
    &&\bH(\tau)\to\tilde\bH(\tau):=\cU_0(\tau,\tau_0)^{-1}\bH(\tau)\cU_0(\tau,\tau_0)-i
    \cU_0(\tau,\tau_0)^{-1}\frac{\partial}{\partial\tau}\cU_0(\tau,\tau_0),
    \label{H-TH}\\
    &&\bU(\tau,\tau_0)\to\tilde\bU(\tau,\tau_0):=\cU_0(\tau,\tau_0)^{-1}\bU(\tau,\tau_0).
    \label{U-TU}
    \eea
It is easy to see that the adiabatic approximation corresponds to $\tilde\bH(\tau)\approx 0$ and $\tilde\bU(\tau,\tau_0)\approx 1$. Expressing $\bU(\tau,\tau_0)$ as a time-ordered exponential of $\tilde\bH(\tau)$ and using (\ref{U-TU}) we obtain the following adiabatic series expansion for the original evolution operator.
    \be
    \bU(\tau,\tau_0)=\cU_0(\tau,\tau_0)\sT e^{-i\int_{\tau_0}^\tau d\tau'\,\tilde\bH(\tau')}=
    \cU_0(\tau,\tau_0)\left[1+\sum_{\ell=1}^\infty \tilde\bU^{(\ell)}(\tau,\tau_0)\right]=
    \sum_{n=0}^\infty\cU_n(\tau,\tau_0),
    \label{ASE}
    \ee
where, for all $\ell\geq 1$,
    \bea
    \tilde\bU^{(\ell)}(\tau,\tau_0)&:=&
    (-i)^\ell\int_{\tau_0}^\tau d\tau_\ell \int_{\tau_0}^{\tau_\ell} d\tau_{\ell-1}\cdots
    \int_{\tau_0}^{\tau_2} d\tau_1 \tilde\bH(\tau_\ell)\tilde\bH(\tau_{\ell-1})\cdots\tilde\bH(\tau_1),
    \label{wq1}
    \\
    \cU_\ell(\tau,\tau_0)&:=&\cU_0(\tau,\tau_0)\tilde\bU^{(\ell)}(\tau,\tau_0).
    \label{wq2}
    \eea

Substituting (\ref{ASE}) in (\ref{S}) we obtain the following adiabatic series expansion for the transfer matrix of $v$.
    \be
    \bM=\bU_0(\tau_+)^{-1}\left[\sum_{n=0}^\infty\cU_n(\tau_+,\tau_-)\right]\bU_0(\tau_-).
    \label{wq0}
    \ee
With the help of (\ref{wq2}) and (\ref{wq0}) we can write this relation in the form
    \be
    \bM=\sum_{n=0}^\infty \bM^{(n)}=\bM^{(0)}\left[1+\sum_{\ell=1}^\infty\bA^{(\ell)}\right],
    \label{bM=}
    \ee
where
    \bea
    \bM^{(0)}&:=&\bM_{\rm sc}:=\bU_0(\tau_+)^{-1}\cU_\ell(\tau_+,\tau_-)\bU_0(\tau_-),~~~~~~~~~~
    \bM^{(\ell)}:=\bM^{(0)}\bA^{(\ell)},
    \label{Ms=:}\\
    \bA^{(\ell)}&:=&\bU_0(\tau_-)^{-1}\tilde \bU^{(\ell)}(\tau_+,\tau_-)\bU_0(\tau_-).
    \label{Ns=:}
    \eea

Keeping the first $n+1$ terms on the right-hand side of (\ref{bM=}) and ignoring others give rise to an $n$-th order improved adiabatic/semiclassical approximation for $\bM$, namely
    \be
    \bM\approx \sum_{\ell=0}^n\bM^{(\ell)}=\bM^{(0)}\left[1+\sum_{\ell=1}^n\bA^{(\ell)}\right].
    \label{n-th-order}
    \ee
The choice $n=0$ leads to the semiclassical expression for $\bM$ that is given in \cite{p115}. The computation of the higher order terms for a finite-range potential with one or both of $\tau_\pm$ taking finite values is quite involved. In the following we examine the structure of (\ref{wq0}) for $\tau_\pm=\pm\infty$, which applies for both finite- as well as infinite-range potentials.

The calculation of $\bM_{\rm sc}$ is straightforward. It results in the following semiclassical expression for the transfer matrix \cite{p115}.
    \be
    \bM\approx\bM_{\rm sc}:=\left[
    \begin{array}{cc}
    e^{i\eta(\infty)}& 0\\
    0 & e^{-i\eta(\infty)}\end{array}\right],
    \label{wkb-infinite}
    \ee
where
    \be
    \eta(\tau):=\int_{-\infty}^\tau[\fn(\tau')-1]\,d\tau'=
    \int_{-\infty}^x[\sqrt{k^2-v(x')}-k]\,dx'.
    \label{eta=}
    \ee

To determine the higher order corrections to (\ref{wkb-infinite}), we need to compute $\tilde\bH(\tau)$. It is not difficult to show (using the results of \cite{jmp-1999} or by direct calculation) that
    \bea
    \tilde\bH(\tau)&=&\frac{i\dot\fn(\tau)}{2\fn(\tau)}\sum_{a=\pm}
    e^{2ai\delta(\tau)}|\Psi_a(\tau_0)\kt\br\Phi_a(\tau_0)|\nn\\
    &=&\frac{i\dot\fn(\tau)}{2\fn(\tau)}\,\Big\{\cos[2\delta(\tau)]\:\sigma_1+
    \sin[2\delta(\tau)]\left(\fa_+\sigma_2-i\fa_-\sigma_3\right)\Big\},
    \label{wq4}
    \eea
where
    \be
    \delta(\tau):=\delta_-(\tau)=\int_{\tau_0}^{\tau}\fn(\tau')d\tau',~~~~~~
    \fa_\pm:=\frac{1}{2}\left[\fn(\tau_0)\pm \fn(\tau_0)^{-1}\right].
    \ee
For $\tau_0=\tau_-=-\infty$, we have $\fn(\tau_0)=\fa_+=1$, $\fa_-=0$, and (\ref{wq4}) reduces to
    \be
    \tilde\bH(\tau)=\frac{i\dot\fn(\tau)}{2\fn(\tau)}\,\left[
    \begin{array}{cc}
    0 & e^{-2i\delta(\tau)}\\
    e^{2i\delta(\tau)} & 0\end{array}\right].
    \label{wq5}
    \ee
The fact that $\tilde\bH(\tau)$ has vanishing diagonal entries implies that $\tilde \bU^{(\ell)}(\tau_+,\tau_-)$ (and consequently $\bA^{(\ell)}$ and $\bM^{(\ell)}$) are diagonal matrices for even $\ell$ and have vanishing diagonal entries for odd $\ell$. Because a $2\times 2$ matrix with vanishing diagonal entries can be written as $\sigma_1$ times a diagonal matrix, we have
    \be
    \bA^{(\ell)}=\sigma_1^{\nu_\ell}\left[\begin{array}{cc}
    A_\ell^- & 0\\
    0 & A_\ell^+ \end{array}\right],
    \label{N-ell=2}
    \ee
where we have used (\ref{wq1}) -- (\ref{Ns=:}) and (\ref{wq5}), and introduced
    \bea
    &&\nu_\ell:=\frac{1-(-1)^\ell}{2}=\left\{\begin{array}{cc}
    1&{\rm~for~odd}~\ell,\\
    0&{\rm~for~even}~\ell,\end{array}\right.\\[6pt]
    &&A_\ell^\pm:=\frac{1}{2^\ell}\int_{-\infty}^\infty d\tau_\ell\int_{-\infty}^\infty d\tau_{\ell-1}\cdots\int_{-\infty}^\infty d\tau_1~ e^{\pm2i\varphi_\ell(\tau_1,\cdots,\tau_\ell)}
    \prod_{j=1}^\ell\frac{\theta(\tau_j-\tau_{j-1})
    \dot\fn(\tau_j)}{\fn(\tau_j)},
    \label{Npm=}\\[6pt]
    &&\varphi_\ell(\tau_1,\cdots,\tau_\ell):=
    \left\{\begin{array}{cc}
    \displaystyle  -\eta(\tau_1)-\tau_1 &{\rm~for}~\ell=1,\\[6pt]
    \displaystyle  -\eta(\tau_\ell)-\tau_\ell+
    \sum_{j=1}^{(\ell-1)/2}\int_{\tau_{2j-1}}^{\tau_{2j}}d\tau\fn(\tau)&{\rm~for~odd}~\ell>1,\\
    \displaystyle \sum_{j=1}^{\ell/2} \int_{\tau_{2j-1}}^{\tau_{2j}}d\tau\fn(\tau)&{\rm~for~even}~\ell.\end{array}\right.
    \label{varphi=}
    \eea

In light of the fact that the integrand on the right-hand side of (\ref{Npm=}) involves $\dot\fn/\fn$, which is the derivative of $\ln\fn$, we can perform integration by parts to simplify this relation. For $\ell=1$ and $2$, this gives
    \bea
    &&A^\pm_1=\pm i\int_{-\infty}^\infty d\tau~e^{\mp 2i[\eta(\tau)+\tau]}\;\fm_1(\tau),
    \label{N1=2}\\
    &&A^\pm_2=\mp\frac{i}{2}\int_{-\infty}^\infty d\tau~\fm_2(\tau)+
    \int_{-\infty}^\infty d\tau_2\int_{-\infty}^{\tau_2} d\tau_1~
    e^{\displaystyle\mbox{$\pm 2i\int_{\tau_1}^{\tau_2}d\tau\fn(\tau)$}}\fm_1(\tau_1)\fm_1(\tau_2),
    \label{N2=2}
    \eea
where we have introduced
    \[\fm_\ell(\tau):=\fn(\tau)\ln[\fn(\tau)]^\ell,\]
and made use of (\ref{N-ell=2}) -- (\ref{varphi=}). It is important to observe that (\ref{N1=2}) and (\ref{N2=2}) do not involve $\dot\fn$.

Because $\ln\!\fn=\!\fn-1+\cO(\!\fn-1)^2$, (\ref{N1=2}) and (\ref{N2=2}) suggest that we can express $A^{\pm}_\ell$ as a sum of integrals whose integrand is of the order of
    \[[\fn(\tau)-1]^{\ell}=\left[\sqrt{1-v(x)/2k^2}-1\right]^{\ell}=
    \cO\left[v(x)/2k^2\right]^{\ell}.\]
Consequently, for an energy-independent potential $v$, (\ref{n-th-order}) is a good approximation for large $k$ (high energies). In optical applications, where $\fn$ corresponds to the refractive index of the medium, this approximation is reliable provided that $\fn$ deviates from unity by small amounts, $|\fn-1|\ll 1$. Because this condition does not restrict $|\dot\fn|$, it is weaker than the adiabaticity (semiclassicality) condition (\ref{adi-condi-0}). This, in particular, implies that we can apply the $n$-th order improved semiclassical approximation (\ref{n-th-order}) with $n\geq 1$ to any finite-range potential fulfilling $|\fn-1|\ll 1$. For a generic finite-range potential, $|\dot\fn|$ can take arbitrarily large values at the boundary points of its support, and the semiclassical approximation (\ref{wkb-infinite}) cannot be applied.

Consider for example a locally periodic refractive index profile of the form \cite{invisible,pra-2014}
    \be
    \fn=\left\{\begin{array}{cc}
    1+\epsilon\,e^{iK x} & {\rm for}~x\in(0,L),\\
    1 & {\rm for}~x\notin(0,L),\end{array}\right.=\left\{\begin{array}{cc}
    1+\epsilon\,e^{iK\tau/k} & {\rm for}~\tau\in(0,kL),\\
    1 & {\rm for}~\tau\notin(0,kL).\end{array}\right.
    \label{exp-pot}
    \ee
where $\epsilon, K,L$ are real parameters, $K$ and $L$ are positive, and $|\epsilon|\ll 1$. We can easily show that
    \bea
    \eta(\tau)&=&\left\{\begin{array}{ccc}
    0 & {\rm for} & \tau\leq 0,\\[6pt]
    \frac{i\epsilon k}{K}\left(1-e^{iK\tau/k}\right) & {\rm for} & 0<\tau<LK,\\[6pt]
    \frac{i\epsilon k}{K}\left(1-e^{iKL}\right)& {\rm for} & \tau\geq LK,
    \end{array}\right.\\
    A_1^\pm &=&\pm \left[\frac{e^{i(K\mp 2k)L}-1}{K\mp 2k}\right]k\epsilon+\cO(\epsilon^2).
    \eea
Substituting these relations in (\ref{N-ell=2}) and making use of (\ref{wkb-infinite}), (\ref{bM=}), and (\ref{M}), we obtain the following expressions for the reflection and transmission amplitudes.
    \bea
    R^r&=&\frac{\left[e^{i(K- 2k)L}-1\right]k\,\epsilon}{K-2k}+\cO(\epsilon^2),\\
    R^r&=&\frac{\left[e^{i(K+ 2k)L}+1\right]k\,\epsilon}{K+2k}+\cO(\epsilon^2),\\
    T&=&1+\frac{\left(e^{iKL}-1\right)k\,\epsilon}{K}+\cO(\epsilon^2).
    \eea
These agree with the perturbative results reported in Ref.~\cite{pra-2014}.

In principle, one can calculate $A^\pm_\ell$ for larger values of $\ell$ (at least numerically) and construct the improved semiclassical approximation~(\ref{n-th-order}) for any scattering potential. For $n\geq 1$, this is a reliable approximation provided that $|\fn-1|$ is small. The same is true for the perturbative expansion of the transfer matrix that we discuss in \cite{pra-2014}. However the improved semiclassical approximation accounts also for certain non-perturbative effects.

In Refs.~\cite{pra-1997a,jmp-1999,book1} we discuss an alternative to the adiabatic series expansion that we call (generalized) adiabatic product expansion. The latter yields the evolution operator of a Hermitian (non-Hermitian) time-dependent Hamiltonian as the product of the adiabatic evolution operators for a sequence of Hamiltonians $\bH_\ell$ that are obtained by successive application of the transformations of the type (\ref{H-TH}). One can obtain exactly solvable cases where $\bH_{\ell_\star}$ vanishes for some $\ell_\star$ and the product expansion terminates. A direct application of this approach for the case that $\bH_0$ is the Hamiltonian (\ref{def2}) fails, because $\bH_0=\bH(\tau)$ implies $\bH_{2\ell-1}=\tilde\bH(\tau)$ and $\bH_{2\ell}=\bH(\tau)$ for all $\ell\geq 1$. One may seek for a variant of this scheme where after a transformation of the type (\ref{H-TH}) one performs another time-dependent linear transformation \cite{jmp-1999}. In general this yields infinite product expansions for $\bU(\tau,\tau_0)$ and $\bM$. The condition that the latter expansion terminates is equivalent to the exact solvability of the scattering problem. Exploring the effectiveness of this approach for finding exactly solvable scattering potentials requires further investigation.

\subsection*{Acknowledgments}  This work has been supported by  the Scientific and Technological Research Council of Turkey (T\"UB\.{I}TAK) in the framework of the project no: 112T951, and by the Turkish Academy of Sciences (T\"UBA).


\ed